\documentclass[11pt,a4paper]{article}

\usepackage{natbib}
\usepackage[english]{babel}
\usepackage[utf8x]{inputenc}
\usepackage[T1]{fontenc}
\usepackage{amssymb}
\usepackage{enumitem}

\usepackage[toc,title,page]{appendix}

\usepackage[a4paper,top=3cm,bottom=3cm,left=2.5cm,right=3.0cm,marginparwidth=1.75cm]{geometry}

\usepackage{amsmath}
\usepackage{graphicx}
\usepackage[colorinlistoftodos]{todonotes}
\usepackage[colorlinks=true, allcolors=blue]{hyperref}
\usepackage{bm}
\usepackage{float}
\usepackage{rotating}
\usepackage{subcaption}

\usepackage[colorinlistoftodos]{todonotes}
\definecolor{darkolivegreen}{rgb}{0.33, 0.42, 0.18}
\definecolor{mediumpersianblue}{rgb}{0.0, 0.4, 0.65}
\definecolor{firebrick}{rgb}{0.7, 0.13, 0.13}

\title{On some extensions of shape-constrained generalized additive modelling in R}
\author{Natalya Pya Arnqvist \\
          \small Department of Mathematics and Mathematical Statistics, \\
         \small Ume\aa{}  University, 901 87 Ume\aa{}, Sweden\\[0.1cm]
         \small natalya.pya@umu.se
          }

\newcommand{\X}{{\bf X}}

\providecommand{\keywords}[1]
{
  \small	
  \textbf{\textit{Keywords:}} #1
}

\begin{document}
\maketitle

\begin{abstract}
   Regression models that incorporate smooth functions of predictor variables to explain the relationships with a response variable have gained widespread usage and proved successful in various applications. 
   By incorporating smooth functions of predictor variables, these models can capture complex relationships between the response and predictors while still allowing for interpretation of the results. In situations where the relationships between a response variable and predictors are explored, it is not uncommon to assume that these relationships adhere to certain shape constraints. Examples of such constraints include monotonicity and convexity.  The \verb+scam+ package for R has become a popular package to carry out the full fitting of exponential family generalized additive modelling with shape restrictions on smooths. The paper aims to extend the existing framework of shape-constrained generalized additive models (SCAM) to accommodate smooth interactions of covariates, linear functionals of shape-constrained smooths and incorporation of residual autocorrelation. The methods described in this paper are implemented in the recent version of the package \verb+scam+, available on the Comprehensive R Archive Network (CRAN).
\end{abstract}


\section{Introduction}
Any application area that analyzes the relationship between a response and multiple covariates could potentially benefit from using nonparametric and semiparametric regression models. 
When analyzing the relationships between a response and predictors, it might be natural to assume that some relationships obey certain shape constraints, such as monotonicity and convexity. Such problems are widespread in ecological and environmental studies. For example, a county's number of protest events is expected to increase with population size. The relationship between tree height and altitude is known to decrease, but tree height increases with tree age. The dose-effect curve in medicine, the relationship between daily mortality and air pollution concentration, body mass index and incidence of heart diseases are other examples where a shape constraint is required. In such studies, unconstrained flexible nonparametric modelling might be too flexible and give implausible or un-interpretable results. Shape-constrained generalized additive models (SCAM), proposed in \cite{pya2015} and implemented in an R package \verb+scam+ \citep{scam24}, provide a general and powerful framework for semiparametric regression modelling with shape constraints. SCAM is used in various application areas, including ecological and environmental studies, 
  education, public health, 
   forestry, 
  genetics, 
 medical and psychological research.

The additive models in the presence of shape constraints have been studied in the larger literature on regression. Construction of shape-constrained spline-based smoothers can be done by using linear inequality 
constraints on the spline coefficients  \citep{villalobos1987,ramsay88,kelly90,zhang04,wang2015,meyer18}. However, the linear inequality constraints bring methodological difficulties in estimating the degree of spline smoothness. 
Several other papers have studied additive shape-constrained regression based on B-splines, including \cite{he98,bollaerts06,rousson08,wang11,wang2021}.
\cite{groeneboom2001,groeneboom2008,guntuboyina2015} has considered the problem of univariate convex regression.
Alternative approaches are based on non-spline ideas. \cite{mammen1991,hall2001,du2013} focused on shape constrained kernel regression techniques, whereas\cite{antoniadis2007} proposed a penalized wavelet regression. 
There was also work in the direction of additive models without smoothing that has been covered by \cite{mammen2007,cheng2009,fang2012,cheng2012,yu2014,chen2016}.
The problem of monotonic additive regression with Bayesian ideas was considered by \cite{holmes03,lang04,dunson03,dunson05}, and with boosting approach by \cite{tutz2007}.
 \cite{brezger2008} developed a Bayesian approach for GAM using penalized B-splines and incorporated monotonicity assumption on univariate smooth by introducing specific prior distributions.    
Despite such diverse literature on shape-constrained regression, existing procedures either lack the ability to efficiently estimate the smoothing parameters in multiple model components setting or result in
non-smooth functions. 

The models with multi-dimensional smooths under shape constraints have been much less studied. \cite{bollaerts06} introduced bivariate shape-constrained P-splines in the least square setting with two covariates only. 
\cite{lin2014} proposed a Bayesian regression method for estimating monotone multi-dimensional functions, building on Gaussian process projection.
\cite{dette2006} developed monotone smooth regression for several covariates based on kernel regression. 
A slightly more general case was considered by \cite{du2013}, who extended the idea of \cite{hall2001} of monotone kernel regression smoothing to a multivariate setting,
covering monotonicity and concavity. Except for the SCAM, the existing approaches work only with normal responses and a non-additive model structure.

The main objective of this paper is to provide an extension of the existing framework for generalized additive modelling with a mixture of unconstrained terms and various shape-restricted terms to accommodate smooth interaction of covariates, varying coefficient terms, linear functionals with or without shape constraints as model components, and data with short-term temporal or spatial autocorrelation.
This paper contributes to the following novel elements: i) bivariate interaction smooths with increasing and decreasing constraints are proposed, which allows for functional ANOVA decomposition (smooth ANOVA) within generalized additive modelling with shape constraints; ii) extension of SCAM with AR1 model on the residuals is introduced, where residual auto-correlation with an AR1 correlation structure can be dealt with in Gaussian models with identity links; iii) extension of SCAM to allow linear functionals of shape-constrained smooths as model components, is proposed, which is known as scalar-on-function regression in the field of functional data analysis.

The remainder of the paper is structured as follows. The general modelling framework is introduced in Section \ref{section:framework}. Section \ref{section:extendedscam} describes the SCAM extensions and illustrates their use in examples. Conclusions and some discussions are given in Section \ref{section:conclusions}.

\section{The general SCAM framework}\label{section:framework}
The class of models considered can be written as  
\begin{equation}\label{eqn:scam}
g(\mu_i) = {\bf A}_i {\bm \theta} + \sum_j f_j(z_{ji}) + \sum_k m_k(x_{ki}),~~~ Y_i \sim {\rm  EF}(\mu_i,\phi), 
\end{equation}
where  $g$ is a specified link function, $Y_i$ is a random real-valued response, for $i=1,\ldots, n,$ with the mean $\mu_i$ with an exponential family distribution (scale parameter $\phi$), ${\bf A}_i$ is 
the $i$th row of a known parametric model matrix with an associated vector of unknown coefficients $\bm \theta,$ $f_j$ is an unknown smooth function of covariate $z_j$, and $m_k$ is an 
unknown {\em shape constrained} smooth function of covariate $x_k$. The covariates $z_j$ and $x_k$ may be vector-valued.

What makes models in (\ref{eqn:scam}) different from the standard generalized additive models (GAM) are the shape constraints on $m_k.$ These models gained popularity through \cite{pya2015}, which
provided computational efficient estimation methods. The work recognizes that the full practical benefits from allowing 
flexible dependence on covariates could only be realized if the degrees of smoothing of the $f_j$ and $m_k$ are estimated as part of model fitting. 
\cite{pya2015} proposed shape-constrained P-splines (SCOP-splines) based on a nonlinear reparametrization of P-splines \citep{eilers1996} with discrete penalties, which allows a variety of shape
constraints for univariate and multivariate smooths. Specifically, for a univariate smooth
$$
  m(x)=\sum_{j=1}^{q}\gamma_jb_j(x),
$$
where $q$ is the number of basis functions, $b_j$ are B-spline basis functions of at least second order, $\gamma_{j}$ are unknown coefficients. Using the first order derivative feature of the B-splines, 
shape constraints are imposed by re-parameterizing the spline coefficients 
$
  \bm\gamma=\bm\Sigma\tilde{\bm\beta},
$
where  $\bm\gamma=(\gamma_{1},\ldots, \gamma_{q})^T,$ 
$\bm\Sigma$ is a matrix of 0's and 1's,
$\bm\beta=(\beta_{1},\ldots,\beta_{q})^T,$ and $\tilde{\bm\beta}$ is a vector of $\bm\beta$ elements, where depending on the shape constraints, some of the elements are exponentiated  $\beta_{j}$.  
For example, in case of a monotonically increasing smooth, $\tilde{\bm\beta}=\left[\beta_1,\exp(\beta_2),\ldots,\exp(\beta_q) \right]^T,$ while $\Sigma_{ij}=0$ if $i<j$ and $\Sigma_{ij}=1$ if $i\geq j.$
The $n$ vector of the evaluated function values, $m(x),$ can now be expressed as 
$
 {\bf m}=\X\bm\Sigma\tilde{\bm\beta},
$
where $\X$ is a matrix with $X_{ij}=b_j(x_i).$ In a smoothing context, with each $m(x)$ is associated a smoothing penalty, which is quadratic in the $\beta_j$ and controls the `wiggliness' of $m.$
In a matrix form, the penalty can be written as $\bm\beta^T{\bf S}\bm\beta,$ where $\bf S$ is a matrix of known coefficients. 
Multivariate shape-constrained smooths with constraints assumed on either all or a selection of the covariates are built up using the concept of tensor product splines.

To represent (\ref{eqn:scam}) for computation, all unconstrained $f_j$ are represented via basis expansions of modest rank, with chosen penalties and identifiability constraints. This allows
to specify $\sum_j f_j(z_{ji})={\bf F}_i\bm\gamma,$ where $\bf F$ is a model matrix, $\bm\gamma$ is a vector of unknown coefficients. The penalties on the $f_j$ are quadratic in $\bm\gamma.$
By absorbing the corresponding identifiability constraints, the model matrices for all the $m_k$ can be combined so that $\sum_km_k(x_{ki})={\bf M}_i\tilde{\bm\beta},$
where $\bf M$ is a model matrix, $\tilde{\bm\beta}$ is a vector containing a mixture of $\beta_j$ and exponentiated $\beta_j.$
So (\ref{eqn:scam}) becomes 
$
  g(\mu_i)={\bf A}_i {\bm \theta}+{\bf F}_i\bm\gamma+{\bf M}_i\tilde{\bm\beta},
$
 and by combining the matrices column-wise into one model matrix $\X=\left[{\bf A}:{\bf F}:{\bf M} \right],$
\begin{equation}\label{eqn:matrixscam}
    g(\mu_i)=\X_i\tilde{\bm\beta}, ~~~ Y_i \sim {\rm  EF}(\mu_i,\phi), 
\end{equation}
where $\tilde{\bm\beta}$ now contains $\bm\theta,$ $\bm\gamma$ and original  $\tilde{\bm\beta}.$
Let $l(\bm\beta)$ be the model log likelihood and ${\bf S}^k$ is the penalty matrix of known coefficients that correspond to the $k$th smoothing penalty.
Generally, ${\bf S}^k$ has only a small block of non-zero elements.
The estimated model coefficients are  then
\begin{equation}\label{eqn:maxlik}
  \hat{\bm\beta}=\underset{\bm\beta}{\textrm{argmax}}\left\{ l(\bm\beta)-\frac{1}{2}\sum_k^M \lambda_k\bm\beta^T{\bf S}^k \bm\beta\right\},
\end{equation}
given $M$ values for the smoothing parameter $\lambda_k,$ controlling the level of penalization.
Smoothing parameters are estimated to minimize a prediction error criterion
such as GCV or AIC in the outer optimization scheme.
Since the imposed extra non-linearity of SCOP-splines coefficients bans re-using or modifying the methods developed for unconstrained GAMs,
different computational schemes for estimating the model coefficients, $\bm\beta,$ and smoothing parameters, $\lambda_k,$ were developed, as well as simulation-free approximate Bayesian confidence intervals for the model terms \cite{pya2015}. Over the last few years, various SCOP-splines have been introduced (upon scam-users' requests) and implemented in the \verb+scam+ package. The univariate single penalty built-in shape-constrained smooth classes include such constraints as monotonicity, concavity, a mixture of monotonicity and concavity,  monotonicity plus the restriction on passing through zero at the right-end point (left-end point) of the covariate range, and positivity constraint. Seventeen bivariate SCOP-splines under monotonicity and/or concavity constraints are suggested, where monotonicity (concavity) may be assumed on only one of the covariates (single monotonicity) or both of them (double monotonicity) are available as the smooth terms of the SCAM. Double or single monotonicity (concavity) is achieved by the corresponding re-parametrization of the bivariate basis coefficients to satisfy the sufficient conditions formulated in terms of the first order differences of the coefficients. Double penalties for the shape-constrained tensor product smooths are obtained from the penalties
of the marginal smooths.

\section{Extended SCAM}\label{section:extendedscam}

To accommodate the extended SCAMs proposed in this paper, the class of models in (\ref{eqn:scam}) can now be written  as 
\begin{equation}\label{eqn:extendedscam}
g(\mu_i) = {\bf A}_i {\bm \theta} + \sum_j L_{ij}f_j + \sum_k L_{ik}m_k,~~~ Y_i \sim {\rm  EF}(\mu_i,\phi), 
\end{equation}
where $L_{ij}$ and $L_{ik}$ are bounded linear functionals, as in (\ref{eqn:scam}) $f_j$ and $m_k$ are unconstrained and shape-constrained smooth functions of some known covariates correspondingly. 
In the standard case $L_{ik}$ (and $L_{ij}$) is simply a functional of evaluation, $L_{ik}m_k=m_k(x_i),$ however for varying coefficient models $L_{ik}m_k=m_k(x_i)z_i,$ for scalar on function regression $L_{ik}m_k=\int s_i(x)m_k(x)dx.$ Also, a linear random effects term ${\bf U}_i{\bf b},$ where ${\bf b}~\sim ~ N({\bf 0},{\bf I}\sigma_u^2)$
and $\bf U$ is a model matrix that can be added to the model structure of (\ref{eqn:extendedscam}), thus extending the SCAM to shape constrained generalized additive mixed models in the same way as it is done for unconstrained GAM \cite{wood2017book}. The following subsections describe the smooth-ANOVA with shape-constrained smooths, shape-constrained additive modelling with simple autocorrelation on the residuals, and linear functional extensions for SCAM. 

\subsection{SCAM with smooth interactions}
Considering models structured with main effects and interactions can be beneficial in certain scenarios, for instance:
$$
 f_1(x) + f_2(z) + m_3(x, z).
$$
Here, $f_1$ and $f_2$ represent smooth `main effects'. $m_3$ is a smooth `interaction' subject to, for example, an increasing constraint with respect to $x$ and the main effects excluded ($f_1$  can also be subject to an increasing constraint). The construction of such a functional ANOVA decomposition of shape-constrained smooths relies on constructing a tensor product basis. This recognizes the equivalence between tensor product basis construction and the method used for interactions in linear models.  Regression modelling with interactions within generalized additive models, where the will is to separate the unconstrained smooth interaction effect of two or more covariates from the main smooth effects (ANOVA decompositions of smooths), was introduced in \cite{wood2017book}. Following that work,  the marginal smooths of a tensor product are subject to identifiability constraints before constructing the tensor product basis. This leads to interaction smooths that do not encompass the corresponding main effects. 

Two tensor product bivariate interaction smooths under shape constraints have been developed and implemented in the R package \verb+scam+, with increasing and decreasing constraints in the first covariates. Both smooth interactions apply the corresponding SCOP identifiability constraints to the first marginal and centering constraints to the second unconstrained marginal. The SCOP identifiability and centering constraints remove the constant functions from the bases of the marginals. By doing so, constant functions are eliminated from the marginal bases, preventing the interaction term from containing the main effects of $f_1$ and $f_2$. This would happen if the marginal bases were multiplied by the constant functions in the other marginal bases. The removal of constraints does not impact penalty terms, given that constant functions lie within the null space of the penalty.  In the \verb+scam+ package,
two shape-constrained smooth interaction terms with increasing and decreasing constraints are implemented. 

\begin{figure}
\begin{center}
\includegraphics[width=.9\linewidth,height=.9\textwidth]{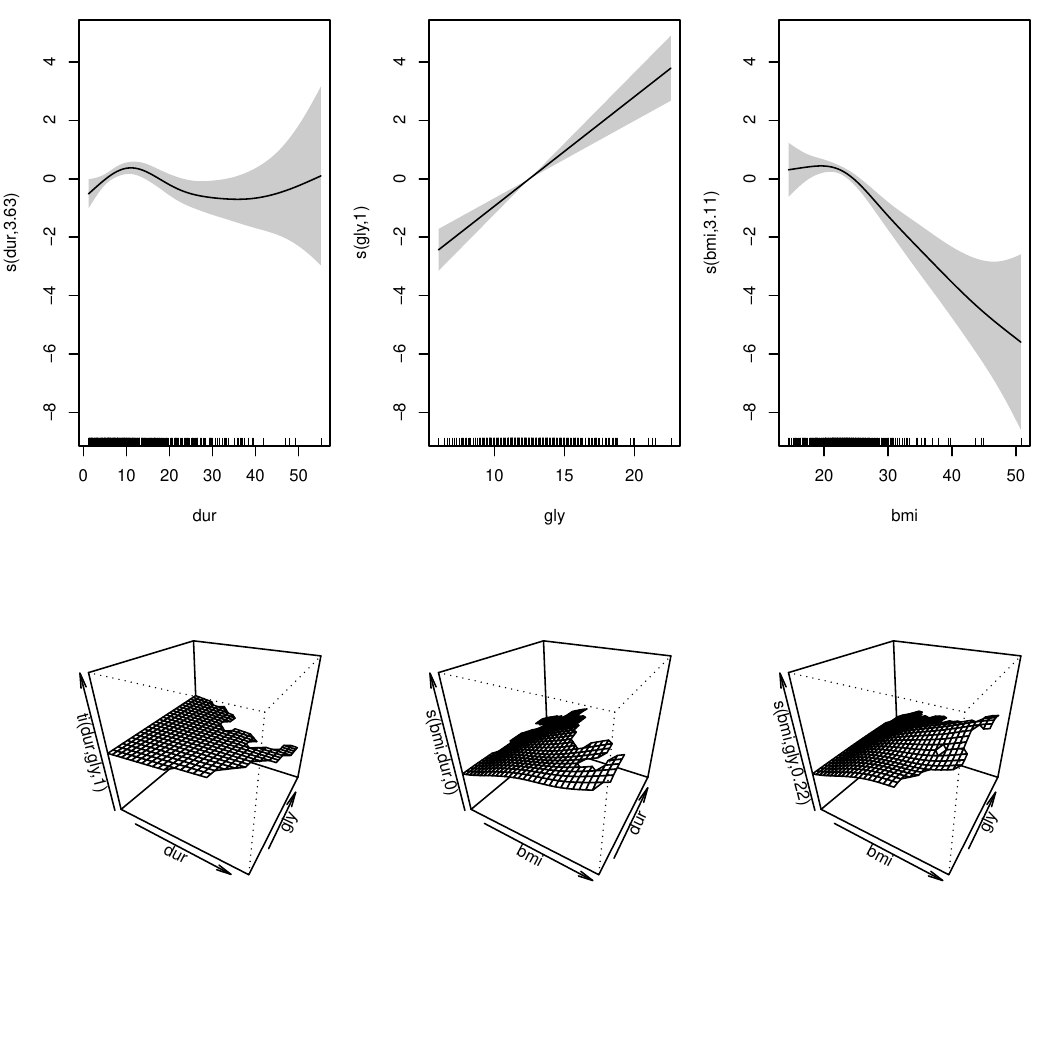}
\end{center}
\caption{Estimated smooth terms of the shape-constrained smooth ANOVA logistic regression model for diabetic retinopathy example.}
\label{fig-ret2}
\end{figure}

As an illustration, consider the \verb+wesdr+ data set from the \verb+gss+ package of Chong Gu. This data set originates from an epidemiological study of diabetic retinopathy. 
It contains 669 observations across four variables: duration of diabetes (\verb+dur+, in years), percent of glycosylated hemoglobin in the blood (\verb+gly+), body mass index (\verb+bmi+), and binary indicator of diabetic retinopathy progression (\verb+ret+). To explore and understand the data, we can start by modelling an unconstrained smooth additive logistic regression model as suggested in \cite{wood2020}. We might expect interactions between covariates, so we include smooth interaction terms. The model structure becomes:
\begin{equation*}
    \begin{split}
        \mathrm{logit}(p_i)= &f_1(\mathtt{dur}_i)+f_2(\mathtt{gly}_i)+f_3(\mathtt{bmi}_i)+ \\
 & f_4(\mathtt{dur}_i,\mathtt{gly}_i)+f_5(\mathtt{dur}_i,\mathtt{bmi}_i)+f_6(\mathtt{gly}_i,\mathtt{bmi}_i),
    \end{split}
\end{equation*}
where $\mathtt{ret}_i ~\sim ~ \mathrm{Binom}(1,p_i),$ $p_i$ is probability of retinopathy, $f_1,f_2,f_3$ are smooth main effects, and $f_4,f_5,f_6$ are smooth interaction terms. The estimated smooths of the unconstrained model shown in figure 7 of \cite{wood2020} indicate a non-zero interaction effect $f_6.$ However, we might expect that if there is such an interaction effect of $\mathtt{gly}_i$ and $\mathtt{bmi}$, the risk of retinopathy will not decrease with increasing body mass index. The unconstrained smooth ANOVA model might be too flexible to catch such an effect. Increasing constraints on the interaction smooths along the $\mathtt{bmi}$ covariate could result in more plausible effects. So, a shape-constrained smooth ANOVA model can be considered here, and it is easily fitted
\begin{verbatim}
m <- scam(ret ~ s(dur,k=k)+ s(gly,k=k)+ s(bmi,k=k)+ ti(dur,gly,k=k)  
         + s(bmi,dur,bs="tismi") + s(bmi,gly,bs="tismi"),
         data=wesdr, family=binomial())     
\end{verbatim}
The estimated smooths are shown in figure \ref{fig-ret2} and reveal sounder results for $f_6$ with the non-decreasing risk effect along the body mass index. 
Introducing shape-constrained interaction smooths leads to only a small increase in AIC. 


\subsection{SCAM with AR1 model on the residuals}
Many environmental and business data sets show short-term temporal, spatial or spatial-temporal autocorrelation. It appears that the fitting problem for non-independent data with a general multivariate normal distribution with an unknown mean and a covariance matrix known up to a constant of proportionality can be transformed to match the fitting problem for independent data with a constant variance exactly. An AR(1) error model can be added to SCAM to deal with the autocorrelation used for the residuals of Gaussian-identity link models in the same way suggested for unconstrained models \cite{wood2015largedata}. The model can be expressed as $Y_i=\X_i\tilde{\bm\beta}+\epsilon_i,$ where $\epsilon_i=\rho\epsilon_{i-1}+e_i$ and $e_i$ are independent $N(0,\sigma^2).$ The covariance matrix for this model, $\mathrm{cov}(\bm\epsilon)=\sigma^2{\bf V}$ is 
\begin{equation*}
     {\bf V}=\left(\begin{matrix}
                  1 & \rho  & \rho^2 & \ldots & \rho^{n-1} \\
                  \rho & 1 & \rho  & \ldots & \rho^{n-2}\\
                  \ldots & \ldots & \ldots & \ldots & \ldots \\
                  \rho^{n-1} & \rho^{n-2} & \rho^{n-3}  & \ldots & 1
               \end{matrix}  
               \right),
\end{equation*}
while the Cholesky factor of the inverse, ${\bf V}^{-1}={\bf C}^T{\bf C}$ is 
\begin{equation*}
     {\bf C}=\frac{1}{\sqrt{1-\rho^2}}\left(\begin{matrix}
                  1 & -\rho  & 0 & \ldots & 0 \\
                  0 & 1 & -\rho  & \ldots & 0  \\
                  \ldots & \ldots & \ldots & \ldots & \ldots \\
                  0  & 0 & 0 & \ldots  &  \sqrt{1-\rho^2}
               \end{matrix}  
               \right),
\end{equation*}
which is banded, having only two non-zero diagonals. So, if we set $\tilde{\bf Y}={\bf CY}$ and $\tilde{\bf X}=\bf CX,$ we have
$
   \tilde{\bf Y} = \tilde{\bf X}\tilde{\bm\beta} + \tilde{\bm\epsilon},
$
where $\tilde{\bm\epsilon}={\bf C}\bm\epsilon$  are now independent $N(0,\sigma^2)$. 
The resulting model is in the form of (\ref{eqn:matrixscam}), so the estimation methods and inference about $\tilde{\bm\beta}$ developed for SCAM with independent data can be used here. The banded structure of the Cholesky factor $\bf C$ makes the transformation by $\bf C$ computationally cheap. Formation of $\tilde{\bf Y}$ and $\tilde{\bf X}$ involves simple weighted differencing operations on the elements of $\bf Y$ and $\bf X$ with $O(n)$ and $O(np)$ operations correspondingly. When $\rho$ is unknown, AIC or GCV can be used to estimate it in a simple one-dimensional direct search. In this case, $\bf C$ should be obtained for every trial $\rho.$ However, as $\bf C$ can be obtained without forming ${\bf V}^{-1}$  and it has the banded structure, the suggested AR(1) error model is computational feasible. If $\rho$ is supplied (as known), then $\bf C$, $\tilde{\bf Y}$ and $\tilde{\bf X}$ are formed only once.

\begin{figure}
\begin{center}
\includegraphics[width=.65\textwidth]{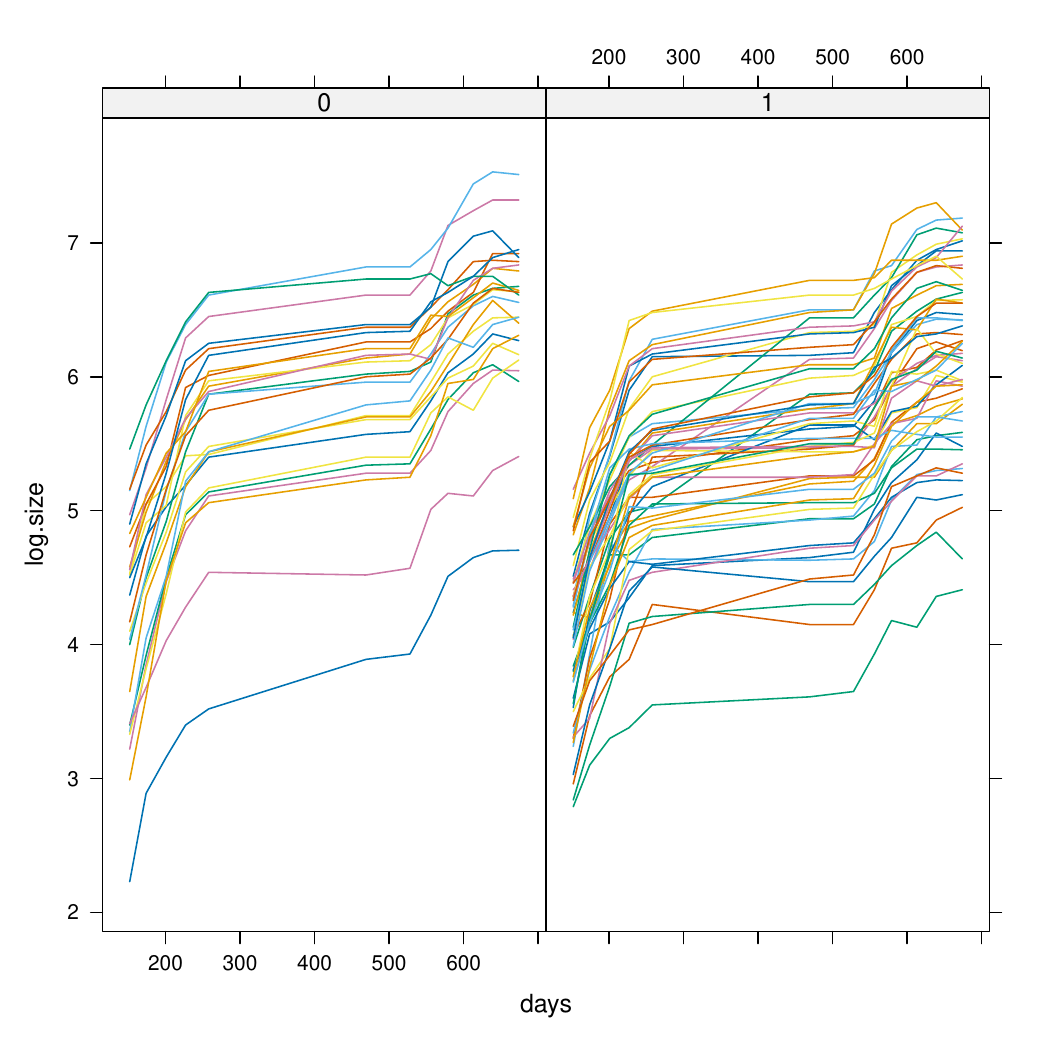}
\end{center}
\caption{Growth paths for 79 Sitka spruce trees: (left) paths for 25 trees grown in control environments; (right) paths for 54 trees grown in ozone-enriched environments.}
\label{fig-sitka1}
\end{figure}

\begin{figure}
\begin{center}
\includegraphics[width=.75\textwidth]{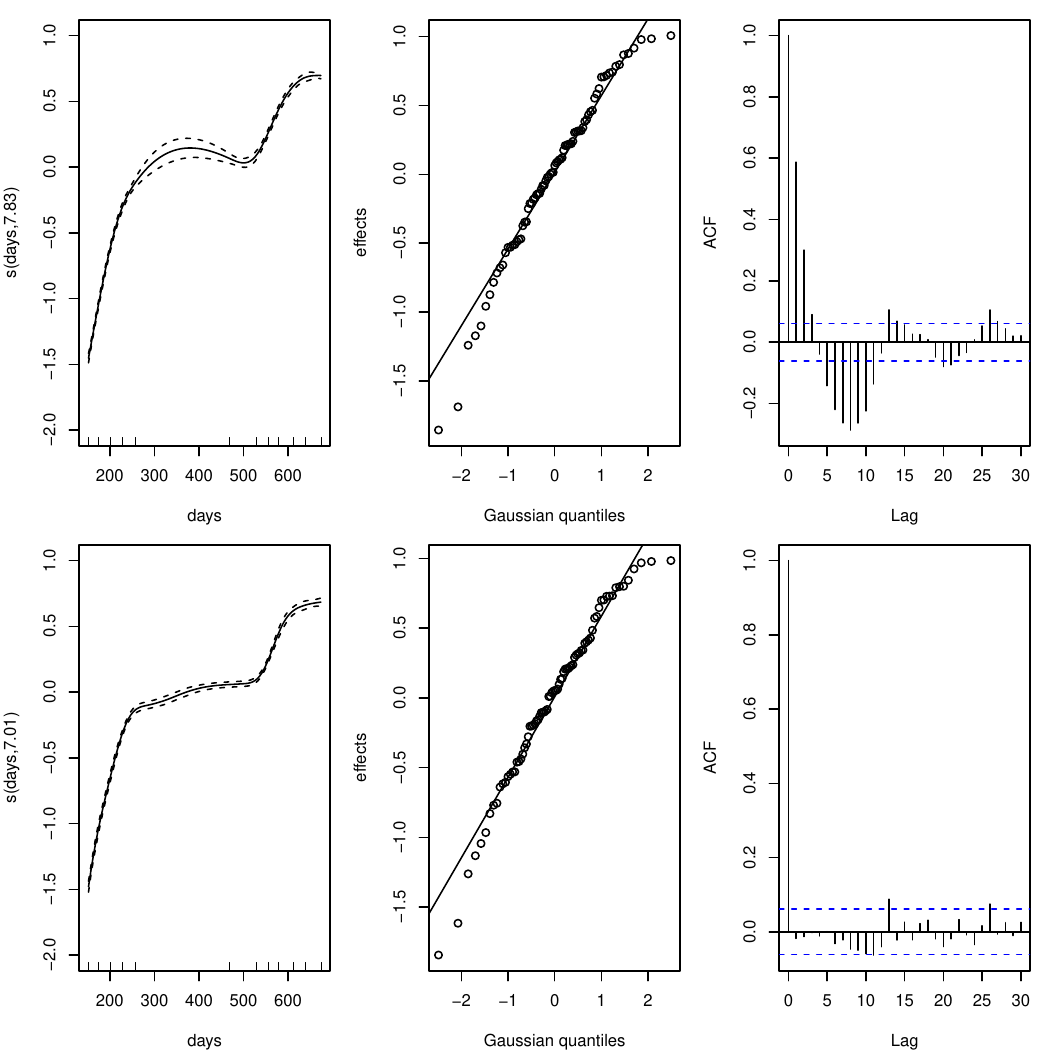}
\end{center}
\caption{The top row is for the unconstrained uncorrelated model while the lower row is for the scam with AR1 error model, s. The first column plots the estimated smooth mean growth curve of Sitka spruce trees with 95\% credible interval. The next column plots the normal QQ-plot for the predicted random effects for the models. The right column shows the ACF plots of the residuals for the un-correlated model (top) and of the standardized residuals for the model s (lower).}
\label{fig-sitka2}
\end{figure}

Consider \verb+sitka+ data set from an R  package \verb+SemiPar+ containing log-size measurements for 79 Sitka spruce trees grown in normal (control) or ozone-enriched conditions. Figure \ref{fig-sitka1} shows the growth trajectories of these trees.
An unconstrained additive mixed modelling for this data (without residual autocorrelation) was illustrated in \cite{wood2017book}. The model estimated smooth mean growth curve reveals a dubious decreasing effect for times in days between 400 and 500 (top left plot in figure \ref{fig-sitka2}), so it is reasonable to impose an increasing constraint on the growth smooth term. In addition, the model residuals contain some autocorrelation (top right plot in figure \ref{fig-sitka2}). So, an AR1 model for the residuals can be included to reduce the effects of the autocorrelation. A simple shape-constrained additive mixed model with an AR1 error model would be 
$$
  \mathtt{log.size}_i= m(\mathtt{day}_i)+\mathtt{ozone}_i+b_{\mathtt{id}_i}+\epsilon_i,
  ~~ b_{\mathtt{id}_i} ~\sim ~N(0,\sigma^2_b),  
$$
where $\epsilon_i$ is an AR1 error model within each tree, $m(\mathtt{day}_i)$ is an increasing smooth term, $\mathtt{ozone}_i$ is a binary factor of ozone treatment, $\mathtt{id}_i$ is the tree identification number to which the $i^{\mathrm{th}}$ observation belongs. The following code would estimate such a model:
\begin{verbatim}
s <- scam(log.size ~ s(days,bs="mpi",k=15)+ozone+s(id.num,bs="re"), 
         data=sitka, AR1.rho=rho, AR.start=start.event)    
\end{verbatim}
Here, \verb+AR.start+ argument is a logical vector indicating where the breaks between autocorrelated sequences of residuals are needed, which is \verb+TRUE+ at the first observation of each tree and \verb+FALSE+ otherwise. \verb+AR1.rho+ is the AR1 correlation parameter, which can be chosen by trying a sequence of values and selecting the value that minimizes the AIC score. To get a likely range of \verb+rho+ values, the autocorrelation of the residuals of the fitted un-correlated model can be used. The lower right plot of figure \ref{fig-sitka2} shows the estimated autocorrelation function of the standardized residuals of the fitted model returned in \verb+s$std.rsd+. The ideal AR1 model would result in standardized residuals that appear approximately uncorrelated. Although not perfect, the model is largely improved compared to the uncorrelated model (top right plot), resulting in a much worse ACF. The fitted model \verb+s+ showed a significant negative effect of the enriched ozone, and figure \ref{fig-sitka2} demonstrates the estimated smooth mean Sitka spruce tree growth curve (lower left) with 95\% credible interval and the normal QQ-plot for the predicted random effect from the model (lower middle). 

\subsection{Linear functionals of shape constrained smooths}
Since the model setup of SCAM is based on the model setup of a standard GAM of the R package \verb+mgcv+ \cite{mgcv}, the methodology developed for SCAM allows for including any unconstrained smooth terms of one or more variables available in the \verb+mgcv+ into its model structure. So, the linear functionals of smooth functions introduced in GAM can be extended to shape-constrained versions. One such example is a scalar on function regression, as known in the field of functional data analysis, which describes the relationship between a scalar response and a set of functional covariates. Suppose that, for subject $i=1,\cdots, n$, we observe data of the form $\left\{y_i,(z_{ij},t_{ij})\right\},~j=1,\cdots, J,$ where $y_i$  is a scalar response and $z_{ij}$  are discrete realizations of a real-valued, square-integrable smooth curve $z_i()$ at observation points $t_{ij}$. A simple functional linear model can be written as 
$$
 Y_i=\alpha +\int m(t)z_i(t)dt +\epsilon_i, ~~\epsilon_i ~\sim ~N(0,\sigma^2),
$$
where $m(t)$ is a smooth function to estimate subject to some shape constraints. Using the SCOP-spline basis expansion $m(t)=\sum_{j=1}^{q}\gamma_jb_j(t),$ the model matrix elements of the linear functional term become $X_{ij}=\int b_j(t)z_i(t)dt.$  Given the linearity of the model in $m(t),$ and by replacing the integral with a discrete sum approximation (numerical quadrature), the model estimation follows the exact methodology developed for SCAMs. The model setup of the \verb+mgcv+ is used to specify the model terms, where a summation convention (summation across rows) is applied when a metric \verb+by+ variable and a smooth covariate are expressed as matrices. 

\begin{figure}
\begin{center}
\includegraphics[width=.75\textwidth,height=.65\textwidth]{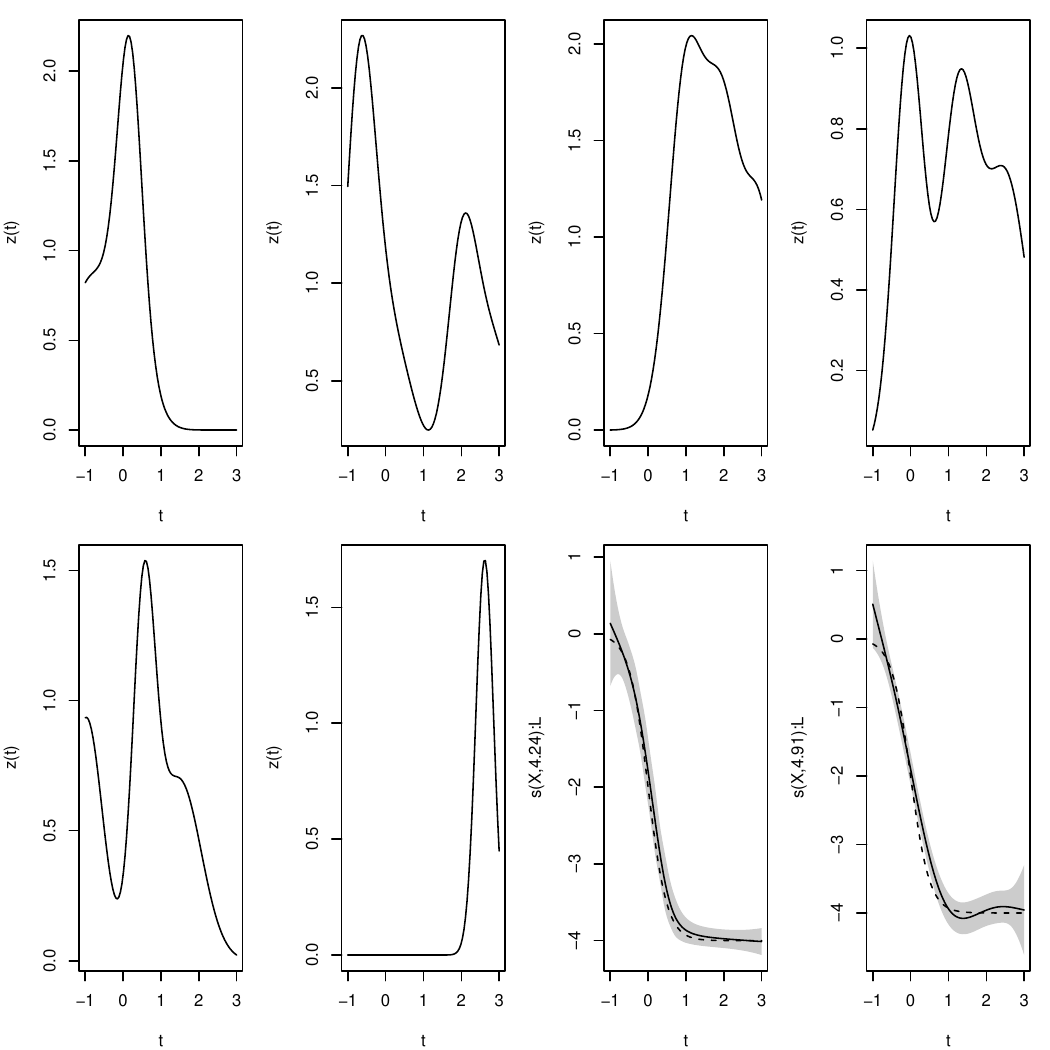}
\end{center}
\caption{Simulated example of the scalar on function regression with decreasing constraint. The first six plots (from top left to lower right) show six examples of the generated functional predictors. In total, 200 functions were simulated and stored in the matrix Z, each at 100 evaluation points. Third lower plot: estimated constrained coefficient function (solid line) and true coefficient function (dashed line). Lower right plot: estimated coefficient function for the unconstrained scalar on function regression (solid line) and true coefficient function (dashed line). }
\label{fig-linfunc}
\end{figure}

To fit the functional linear model with, for example, decreasing constraint, the \verb+scam+ call would be
\begin{verbatim}
b <- scam(y ~ s(X,by=Z,bs="mpdBy"))     
\end{verbatim}
The linear functional term is specified via \verb+s()+ smooth with a \verb+by+ variable, where \verb+Z+ and \verb+X+ are matrices of the same dimensions, \verb+Z+ is a matrix of discrete observations of the smooth functional predictors, \verb+X+ is a matrix of the observation points, $t_{ij}.$ The type of shape constraints imposed on the smooth coefficient function, $m(t)$, is specified in the \verb+bs+ argument. \verb+bs="mpdBy"+ in the example above indicates that $m(t)$ is subject to decreasing constraint.   Eight different SCOP-splines with monotonicity, concavity and a mixture of monotonicity and concavity constraints are implemented in the \verb+scam+ package to be used with a numeric `by' variable that takes more than one value. The construction of these smooth terms is similar to that of the ordinary SCOP-splines, with the only difference being that they are built without applying identifiability constraints, since 
smooth terms with a numeric `by' variable taking more than one value are fully identifiable. 
Figure \ref{fig-linfunc} shows a simulated example of the linear functionals of the smooth function under decreasing constraint. From the two lower right plots, it is clear that the shape-constrained model performs better than the unconstrained version. 

Note that function on scalar regression is also possible with SCAM by expanding the functional coefficients with unconstrained or SCOP-spline basis expansion.
$$
Y_i(t)=f_0(t) + \sum\limits_j x_{ij}f_j(t)+\sum\limits_k x_{ik}m_k(t) + \epsilon_i(t),
$$
where $Y_i$ are functional responses, $x_i$ are scalar predictors, $f_j$ and $m_k$  are functional coefficients to estimate correspondingly without or with shape constraints. Errors, $\epsilon_i(t)$, correlated within a subject can be modelled as AR1 error process with \verb+scam+.

\subsection{Smoothing parameter estimation}
The current fitting approach for SCAM is reasonably efficient and robust when applied to data sets containing a few tens of thousands of observations.
If $n$ and $p$ denote numbers of rows and columns of the model matrix correspondingly, then the approach has the $O(np^2)$ computational cost of standard penalized regression spline based GAM estimation and typically involves 2-4 times as many $O(np^2)$ steps due to the additional non-linearities required for dealing with the shape constrained terms.
As an alternative to the conventional quasi-Newton approach for smoothing parameter $\bm\lambda$ optimization, an extended Fellner-Schall method (EFS) reported in \cite{wood2017fellner} was introduced. The smoothing parameters are estimated in the outer optimization scheme using a full Newton method for coefficient estimation. The extended Fellner-Schall method generalizes the technique of \cite{fellner1986} and \cite{schall1991} and offers a significant simplification over existing methods by re-using quantities anyway required for
model coefficients estimation when iteratively updating $\bm\lambda$.
When modelling with SCAM, rather than maximizing the log restricted marginal likelihood as it is proposed in \citep{wood2017fellner}, similar convenient updates are obtained for minimizing a prediction error GCV criterion, $\mathcal{V}_g=nD(\hat{\bm\beta})/(n-\gamma\tau)^2,$ when the scale parameter of the exponential family distribution is unknown and for the UBRE, $\mathcal{V}_a= D(\hat{\bm\beta})+2\phi\tau\gamma,$ for known scale parameter.
The smoothing parameter update for the GCV minimization is
$$
  \rho_j^*=\rho_j +\log\left\{-\frac{2\gamma D}{n-\tau}\cdot\frac{\partial\tau}{\partial\rho_j}\left[\left(\frac{\partial D}{\partial\bm\beta}\right)^T\cdot\frac{d\hat{\bm\beta}}{d\rho_j} \right]^{-1} \right\},
$$
where $\rho_j=\log(\lambda_j)$ to ensure that $\lambda_j^*$ is positive,
$D$ is the model deviance, $\tau$ is the effective degrees of freedom, $\gamma$ is a parameter that usually has the value $1,$ but can be increased to get smoother models. The $\rho$ update when minimizing the UBRE score is
$$
  \rho_j^*=\rho_j +\log\left\{-2\phi\gamma\cdot\frac{\partial\tau}{\partial\rho_j}\left[\left(\frac{\partial D}{\partial\bm\beta}\right)^T\cdot\frac{d\hat{\bm\beta}}{d\rho_j} \right]^{-1} \right\}.
$$
The EFS approach eliminates the need for implicit differentiation in quasi-Newton optimization and leads to a more robust smoothing parameter scheme for SCAM. It is implemented in the \verb+scam+ package and is called for by  \verb+optimizer="efs"+ argument of the \verb+scam()+ function.

\section{Discussion}\label{section:conclusions}
To date, all developed methods within shape-constrained regression modelling are limited to either modelling with only independent normal response variables or models with standard additive model structures of smooth functions. The main contribution of this work has been to provide practical extensions to the SCAM framework for fitting models that include shape-constrained smooth interactions and linear functionals of the shape-constrained smooths as model components and can deal with residual autocorrelation for the Gaussian-identity link models. The proposed methods seamlessly integrate into the existing general framework of shape-constrained generalized additive modelling, SCAM. So, anyone familiar with generalized additive modelling and shape-constrained additive modelling through R packages such \verb+mgcv+ and \verb+scam+ can readily apply these extensions. 
The methods are implemented in the R package \verb+scam+, which is under continual development and maintenance.

The implementation cost of the suggested EFS method for estimating smoothing parameters is significantly reduced, thus allowing for easier application and opening doors for a wider range of models with shape-constrained terms. The costs can be further reduced by applying a quasi-Newton method for model coefficients estimation in which the Hessian of the log-likelihood will be replaced in the update by a quasi-Newton approximation, thus only requiring the gradient of the log-likelihood. This approach is added as an alternative fitting method (working with the EFS for smoothing parameter selection) in the package \verb+scam+ from version 1.2-14. However, further careful development of the stabilization strategies and convergence investigation are intended. 

In practical terms, \verb+scam+ allows full fitting of exponential family GAMs with shape restrictions on smooths, automatically estimating smoothing parameters and uncertainty estimates. It provides a framework for generalized additive modelling with a mixture of unconstrained terms and various shape-restricted terms of both univariate and bivariate type (bivariate smooths can be subject to shape constraints on either both or one of the covariates), which can also accommodate parametric terms, simple random effects, varying coefficient terms, the interaction of covariates, and also linear functionals with or without shape constraints as model components.

\section*{Acknowledgements}
This work is partly supported by the Swedish Research Council (Reg. No. 2022-04190).

\bibliography{bib-ecosta}

\begin{thebibliography}{42}
\providecommand{\natexlab}[1]{#1}
\providecommand{\url}[1]{\texttt{#1}}
\expandafter\ifx\csname urlstyle\endcsname\relax
  \providecommand{\doi}[1]{doi: #1}\else
  \providecommand{\doi}{doi: \begingroup \urlstyle{rm}\Url}\fi

\bibitem[Antoniadis et~al.(2007)Antoniadis, Bigot, and Gijbels]{antoniadis2007}
A.~Antoniadis, J.~Bigot, and I.~Gijbels.
\newblock Penalized wavelet monotone regression.
\newblock \emph{Statistics \& probability letters}, 77\penalty0 (16):\penalty0
  1608--1621, 2007.

\bibitem[Bollaerts et~al.(2006)Bollaerts, Eilers, and van
  Mechelen]{bollaerts06}
K.~Bollaerts, P.~Eilers, and I.~van Mechelen.
\newblock Simple and multiple {P}-splines regression with shape constraints.
\newblock \emph{British Journal of Mathematical and Statistical Psychology},
  59:\penalty0 451--469, 2006.

\bibitem[Brezger and Steiner(2008)]{brezger2008}
A.~Brezger and W.~J. Steiner.
\newblock Monotonic regression based on {B}ayesian {P}-splines: {A}n
  application to estimating price response functions from store-level scanner
  data.
\newblock \emph{Journal of business \& economic statistics}, 26\penalty0
  (1):\penalty0 90--104, 2008.

\bibitem[Chen and Samworth(2016)]{chen2016}
Y.~Chen and R.~J. Samworth.
\newblock Generalized additive and index models with shape constraints.
\newblock \emph{JRSS (B)}, 78\penalty0 (4):\penalty0 729--754, 2016.

\bibitem[Cheng(2009)]{cheng2009}
G.~Cheng.
\newblock Semiparametric additive isotonic regression.
\newblock \emph{J Stat Plan Inference}, 139\penalty0 (6):\penalty0 1980--1991,
  2009.

\bibitem[Cheng et~al.(2012)Cheng, Zhao, and Li]{cheng2012}
G.~Cheng, Y.~Zhao, and B.~Li.
\newblock Empirical likelihood inferences for the semiparametric additive
  isotonic regression.
\newblock \emph{Journal of Multivariate Analysis}, 112:\penalty0 172--182,
  2012.

\bibitem[Dette and Scheder(2006)]{dette2006}
H.~Dette and R.~Scheder.
\newblock Strictly monotone and smooth nonparametric regression for two or more
  variables.
\newblock \emph{Canadian Journal of Statistics}, 34\penalty0 (4):\penalty0
  535--561, 2006.

\bibitem[Du et~al.(2013)Du, Parmeter, and Racine]{du2013}
P.~Du, C.~F. Parmeter, and J.~S. Racine.
\newblock Nonparametric kernel regression with multiple predictors and multiple
  shape constraints.
\newblock \emph{Statistica Sinica}, pages 1347--1371, 2013.

\bibitem[Dunson(2005)]{dunson05}
D.~Dunson.
\newblock Bayesian semiparametric isotonic regression for count data.
\newblock \emph{JASA}, 100\penalty0 (470):\penalty0 618--627, 2005.

\bibitem[Dunson and Neelon(2003)]{dunson03}
D.~Dunson and B.~Neelon.
\newblock Bayesian inference on order-constrained parameters in glm.
\newblock \emph{Biometrics}, 59:\penalty0 286--295, 2003.

\bibitem[Eilers and Marx(1996)]{eilers1996}
P.~H. Eilers and B.~D. Marx.
\newblock Flexible smoothing with b-splines and penalties.
\newblock \emph{Statistical science}, 11\penalty0 (2):\penalty0 89--121, 1996.

\bibitem[Fang and Meinshausen(2012)]{fang2012}
Z.~Fang and N.~Meinshausen.
\newblock Lasso isotone for high-dimensional additive isotonic regression.
\newblock \emph{J Comput Graph Stat}, 21\penalty0 (1):\penalty0 72--91, 2012.

\bibitem[Fellner(1986)]{fellner1986}
W.~H. Fellner.
\newblock Robust estimation of variance components.
\newblock \emph{Technometrics}, 28\penalty0 (1):\penalty0 51--60, 1986.

\bibitem[Groeneboom et~al.(2001)Groeneboom, Jongbloed, and
  Wellner]{groeneboom2001}
P.~Groeneboom, G.~Jongbloed, and J.~A. Wellner.
\newblock Estimation of a convex function: characterizations and asymptotic
  theory.
\newblock \emph{The Annals of Statistics}, 29\penalty0 (6):\penalty0
  1653--1698, 2001.

\bibitem[Groeneboom et~al.(2008)Groeneboom, Jongbloed, and
  Wellner]{groeneboom2008}
P.~Groeneboom, G.~Jongbloed, and J.~A. Wellner.
\newblock The support reduction algorithm for computing non-parametric function
  estimates in mixture models.
\newblock \emph{Scandinavian Journal of Statistics}, 35\penalty0 (3):\penalty0
  385--399, 2008.

\bibitem[Guntuboyina and Sen(2015)]{guntuboyina2015}
A.~Guntuboyina and B.~Sen.
\newblock Global risk bounds and adaptation in univariate convex regression.
\newblock \emph{Probability Theory and Related Fields}, 163\penalty0
  (1):\penalty0 379--411, 2015.

\bibitem[Hall and Huang(2001)]{hall2001}
P.~Hall and L.-S. Huang.
\newblock Nonparametric kernel regression subject to monotonicity constraints.
\newblock \emph{The Annals of Statistics}, 29\penalty0 (3):\penalty0 624--647,
  2001.

\bibitem[He and Shi(1998)]{he98}
X.~He and P.~Shi.
\newblock Monotone {B}-spline smoothing.
\newblock \emph{JASA}, 93\penalty0 (442):\penalty0 643--650, 1998.

\bibitem[Holmes and Heard(2003)]{holmes03}
C.~Holmes and N.~Heard.
\newblock Generalized monotonic regression using random change points.
\newblock \emph{Statistics in Medicine}, 22:\penalty0 623--638, 2003.

\bibitem[Kelly and Rice(1990)]{kelly90}
C.~Kelly and J.~Rice.
\newblock Monotone smoothing with application to dose-response curves and the
  assessment of synergism.
\newblock \emph{Biometrics}, 46:\penalty0 1071--1085, 1990.

\bibitem[Lang and Brezger(2004)]{lang04}
S.~Lang and A.~Brezger.
\newblock Bayesian {P}-splines.
\newblock \emph{J Comput Graph Stat}, 13\penalty0 (1):\penalty0 183--212, 2004.

\bibitem[Lin and Dunson(2014)]{lin2014}
L.~Lin and D.~B. Dunson.
\newblock Bayesian monotone regression using gaussian process projection.
\newblock \emph{Biometrika}, 101\penalty0 (2):\penalty0 303--317, 2014.

\bibitem[Mammen(1991)]{mammen1991}
E.~Mammen.
\newblock Estimating a smooth monotone regression function.
\newblock \emph{The Annals of Statistics}, pages 724--740, 1991.

\bibitem[Mammen and Yu(2007)]{mammen2007}
E.~Mammen and K.~Yu.
\newblock Additive isotone regression.
\newblock In \emph{Asymptotics: particles, processes and inverse problems},
  pages 179--195. Institute of Mathematical Statistics, 2007.

\bibitem[Meyer(2018)]{meyer18}
M.~C. Meyer.
\newblock A framework for estimation and inference in generalized additive
  models with shape and order restrictions.
\newblock \emph{Statistical Science}, 33\penalty0 (4):\penalty0 595--614, 2018.

\bibitem[Pya(2024)]{scam24}
N.~Pya.
\newblock \emph{scam: Shape Constrained Additive Models}, 2024.
\newblock URL \url{https://CRAN.R-project.org/package=scam}.
\newblock ~R package version 1.2-16.

\bibitem[Pya and Wood(2015)]{pya2015}
N.~Pya and S.~N. Wood.
\newblock Shape constrained additive models.
\newblock \emph{Statistics and computing}, 25:\penalty0 543--559, 2015.

\bibitem[Ramsay(1988)]{ramsay88}
J.~Ramsay.
\newblock Monotone regression splines in action.
\newblock \emph{Stats. Science}, 3\penalty0 (4):\penalty0 425--461, 1988.

\bibitem[Rousson(2008)]{rousson08}
V.~Rousson.
\newblock Monotone fitting for developmental variables.
\newblock \emph{Journal of Applied Statistics}, 35\penalty0 (6):\penalty0
  659--670, 2008.

\bibitem[Schall(1991)]{schall1991}
R.~Schall.
\newblock Estimation in generalized linear models with random effects.
\newblock \emph{Biometrika}, 78\penalty0 (4):\penalty0 719--727, 1991.

\bibitem[Tutz and Leitenstorfer(2007)]{tutz2007}
G.~Tutz and F.~Leitenstorfer.
\newblock Generalized smooth monotonic regression in additive modeling.
\newblock \emph{J Comput Graph Stat}, 16\penalty0 (1):\penalty0 165--188, 2007.

\bibitem[Villalobos and Wahba(1987)]{villalobos1987}
M.~Villalobos and G.~Wahba.
\newblock Inequality-constrained multivariate smoothing splines with
  application to the estimation of posterior probabilities.
\newblock \emph{JASA}, 82\penalty0 (397):\penalty0 239--248, 1987.

\bibitem[Wang and Meyer(2011)]{wang11}
J.~Wang and M.~Meyer.
\newblock Testing the monotonicity or convexity of a function using regression
  splines.
\newblock \emph{The Canadian Journal of Statistics}, 39\penalty0 (1):\penalty0
  89--107, 2011.

\bibitem[Wang and Xue(2015)]{wang2015}
L.~Wang and L.~Xue.
\newblock Constrained polynomial spline estimation of monotone additive models.
\newblock \emph{J Stat Plan Inference}, 167:\penalty0 27--40, 2015.

\bibitem[Wang and Zhou(2021)]{wang2021}
L.~Wang and X.-H. Zhou.
\newblock Estimation of shape constrained additive models with missing response
  at random.
\newblock \emph{Journal of Nonparametric Statistics}, 33\penalty0 (1):\penalty0
  118--133, 2021.

\bibitem[Wood(2023)]{mgcv}
S.~Wood.
\newblock \emph{mgcv: Mixed GAM Computation Vehicle with Automatic Smoothness
  Estimation}, 2023.
\newblock URL \url{https://CRAN.R-project.org/package=mgcv}.
\newblock R package version 1.9.1.

\bibitem[Wood(2017)]{wood2017book}
S.~N. Wood.
\newblock \emph{Generalized Additive Models: An Introduction with {R}}.
\newblock CRC press, 2017.

\bibitem[Wood(2020)]{wood2020}
S.~N. Wood.
\newblock Inference and computation with generalized additive models and their
  extensions.
\newblock \emph{Test}, 29\penalty0 (2):\penalty0 307--339, 2020.

\bibitem[Wood and Fasiolo(2017)]{wood2017fellner}
S.~N. Wood and M.~Fasiolo.
\newblock A generalized fellner-schall method for smoothing parameter
  optimization with application to tweedie location, scale and shape models.
\newblock \emph{Biometrics}, 73\penalty0 (4):\penalty0 1071--1081, 2017.

\bibitem[Wood et~al.(2015)Wood, Goude, and Shaw]{wood2015largedata}
S.~N. Wood, Y.~Goude, and S.~Shaw.
\newblock Generalized additive models for large data sets.
\newblock \emph{JRSS (C)}, 64\penalty0 (1):\penalty0 139--155, 2015.

\bibitem[Yu(2014)]{yu2014}
K.~Yu.
\newblock On partial linear additive isotonic regression.
\newblock \emph{Journal of the Korean Statistical Society}, 43\penalty0
  (1):\penalty0 11--17, 2014.

\bibitem[Zhang(2004)]{zhang04}
J.~Zhang.
\newblock A simple and efficient monotone smoother using smoothing splines.
\newblock \emph{Journal of Nonparametric Statistics}, 16\penalty0 (5):\penalty0
  779--796, 2004.

\end{thebibliography}

\end{document}